\documentclass[a4paper,11pt]{article}
\usepackage{pos}
\usepackage{float}
\usepackage[caption=false]{subfig}
\usepackage{tensor}

\begin{document}

\title{Time-loops in Dirac materials, torsion and unconventional Supersymmetry}
\ShortTitle{Time-loops in Dirac materials, torsion and USUSY}

\author[a]{Alfredo Iorio}
\author*[a,b,c]{Pablo Pais}

\affiliation[a]{Institute of Particle and Nuclear Physics, Charles University,\\ V Hole\v{s}ovi\v{c}k\'{a}ch 2, 18000 Prague 8, Czech Republic}

\affiliation[b]{Instituto de F\'{\i}sica, Facultad de Ciencias, Universidad de la Rep\'{u}blica, \\ Igu\'{a} 4225, Montevideo, Uruguay}

\affiliation[c]{Instituto de Ciencias F\'isicas y Matem\'aticas, Universidad Austral de Chile, \\ Casilla 567, 5090000 Valdivia, Chile}

\emailAdd{iorio@ipnp.troja.mff.cuni.cz}
\emailAdd{pais@ipnp.troja.mff.cuni.cz}

\abstract{We propose a scenario where the effects of dislocations, in bidimensional Dirac materials at low energies, can be described within a Dirac field theory by a vertex proportional to the totally antisymmetric component of the torsion generated by such dislocations. The well-known geometrical obstruction to have a nonzero torsion term of that kind in this two-dimensional settings is overcome through exotic time-loops, obtained from ingeniously manipulated particle-hole dynamics. If such torsion/dislocation is indeed present, a net flow of particles-antiparticles (holes) can be inferred and possibly measured. Finally, we comment on how these discoveries pave the way to a laboratory realization on Dirac materials of Unconventional Supersymmetry, as a top-down description of the $\pi$-electrons in backgrounds with a nonzero torsion.}

\FullConference{%
  40th International Conference on High Energy physics - ICHEP2020\\
  July 28 - August 6, 2020\\
  Prague, Czech Republic (virtual meeting)
}


\maketitle

\section{Torsion and time loops in two-dimensional Dirac materials}
\label{section_introduction}


In General Relativity (GR) curvature is related to the energy content of the spacetime. In alternative theories, torsion, instead, is intimately related to spin \cite{KibbleTorsion1961}. In Einstein-Cartan theory, which is somehow the minimal extension of GR to include torsion, the very existence of spinors produce a torsion field. While other theories as Supersymmetry (SUSY) in curved space, that is Supergravity (SUGRA), as well as the more recent unconventional SUSY (USUSY) \cite{AVZ}, make extensive use of torsion. In condensed matter, the unavoidable existence of defects in crystals, such as disclinations and dislocations, makes natural to include both curvature and torsion in the geometric, continuous description of the elastic properties of materials \cite{Katanaev2005}.

In recent years, due to the structure of their low energy spectrum, Dirac materials \cite{WehlingDiracMaterials2014} have emerged as experimental playgrounds where the fundamental research and the condensed matter one, met. In particular, the role of disclinations is under intense investigation, to realize analogs of Dirac quantum fields in curved spacetimes, see, e.g., \cite{IORIO20111334,Iorio:2011yz}.


If we were able to explore this possibility in two-dimensional Dirac materials, it would be an invaluable help to shed light on some of the mysteries on torsion. One important case is USUSY, especially in its $(2+1)-$dimensional formulation. Indeed, such theory has been found to have many similarities with the Dirac field theory on graphene (see \cite{SU2Ususy,GPZ}, and especially the recent \cite{DauriaZanelli2019}). However, the exploration of the role of torsion in this setting found a geometric obstacle, just due to the $(2+1)$ dimensions: As we shall recall later, a Dirac spinor only couples to the fully antisymmetric component of torsion, hence three dimensions are necessary. Lacking the spatial third dimension, this seemed impossible \cite{deJuan2010}. This ``no-go'' result stopped research in this direction. It was the main goal of the work \cite{original_paper} to suggest a way to surmount this obstacle, based on the use of time as the necessary third dimension. Here we spot the main ideas of this paper.


As well known \cite{WehlingDiracMaterials2014}, the low energy excitations of the $\pi$ electrons of two dimensional Dirac materials with hexagonal lattice, such as graphene, germanene, silicene, are well described by a relativistic-like $(2+1)$-dimensional Dirac theory, governed by the action
\begin{equation}\label{flatAction}
S_0 [\overline{\Psi}, \Psi] = i \hbar v_F \int d^{3}x  \overline{\Psi}\gamma^{a}\partial_{a}\Psi \,,
\end{equation}
where $v_F$ is the Fermi velocity, the flat index $a=0,1,2$, we are in a reducible representation of the Lorentz group, that is $\Psi^T =\left(\psi_{+}, \psi_{-}\right)$ is a \textit{four}-Dirac spinor, made of two irreducible \textit{two}-spinors, $\psi_{\pm}$, describing both Dirac points, and we used the prescription of \cite{IORIO2018265} for the Dirac matrices.

The natural generalization of (\ref{flatAction}) to a $(2+1)$-dimensional spacetime, equipped with a metric $g_{\mu\nu}=\eta_{ab}e^{a}_{\mu}e^{b}_{\nu}$ (being $e^{a}_{\mu}$ the vielbein) and a metric-compatible connection $\Gamma^{\lambda}_{\mu \nu}$ \textit{that includes torsion} \cite{Nakahara} $T^{\lambda}_{\mu \nu} = \Gamma^{\lambda}_{\mu \nu} - \Gamma^{\lambda}_{\nu \mu}$ is
$S = i \hbar v_F  \int d^{3}x \sqrt{-g} \overline{\Psi}\gamma^{\mu} D_{\mu}\Psi$, where the covariant derivative is defined as $D_{\mu}\Psi=\partial_{\mu}\Psi+\frac{i}{2}\omega^{ab}_{\mu}\mathbb{J}_{ab}\Psi$, with $\mathbb{J}_{ab}=\frac{i}{4}[\gamma_{a},\gamma_{b}]$ the Lorentz generators in spinor space. The spin-connection, $\omega^{ab}_{\mu}=e^{a}_{\lambda}(\delta^{\lambda}_{\nu}\partial_{\mu}+\Gamma^{\lambda}_{\mu\nu})e^{b\nu}$, can be decomposed into torsion-free and {\it contorsion} contributions, $\omega_\mu^{ab}=\mathring{\omega}_\mu^{ab}+ \kappa_\mu^{ab}$, where $T^{\lambda}_{\mu\nu}=E_a^{\lambda}
{\kappa_\nu}^{a}_{b} e^{b}_{\mu} - E_a^{\lambda} {\kappa_\mu}^{a}_{b} e^{b}_{\nu}$. Standard manipulations of the action $S$, reported in detail in \cite{original_paper}, lead to the form, apart from possible boundary terms,
\begin{equation}\label{action_torsion}
S = i \hbar v_F\int d^{3}x \; |e| \; \overline{\psi}\left(\gamma^{\mu}\mathring{D}_{\mu} - \frac{i}{4} \gamma^{5} \frac{\epsilon^{\mu\nu\rho}}{|e|} T_{\mu \nu\rho} \right)\psi\;,
\end{equation}
where $|e|=\sqrt{|g|}$, the covariant derivative, $\mathring{D}_{\mu}$, is based on the torsion-free connection, $\mathring{\omega}_\mu^{ab}$, only, $\gamma^{5} \equiv  i \gamma^{0} \gamma^{1} \gamma^{2}$, due to the reducibility of the representation, {\it commutes} with the other gamma matrices, and the contribution due to the torsion is all in the last term through its totally antisymmetric component \cite{Shapiro}. From here, it is evident that the emergent fermions of Dirac materials $\Psi$ can only be coupled to
\begin{equation}\label{T012}
T_{012} \, \mbox{ or} \, T_{102} \, \mbox{ or} \, T_{210} \,,
\end{equation}
hence, one should make sense of the {\ it time component}. In fact, the torsion tensor in crystals is related to the Burgers vector through the formula \cite{Katanaev2005}
\begin{equation}\label{torsion-Burgers}
b^{a}=\int\int_{\Sigma} e^{a}_{\lambda}T^{\lambda}_{\mu \nu}dx^{\mu} \wedge dx^{\nu} \;,
\end{equation}
where $\Sigma$ is a surface containing the dislocation, but otherwise arbitrary, $a = 0,1,2$, and $\wedge$ stands for the exterior product operator. We clearly see that the only two possibilities that a nonzero Burgers vector can give rise to $\epsilon^{\mu\nu\rho}T_{\mu \nu\rho}\neq0$, necessary for the coupling in (\ref{action_torsion}), are: (i) a \textit{time directed} screw dislocation, i.e. $b_{t} \propto \int\int  T_{012} dx \wedge dy$ or (ii) an edge dislocation spotted by a \textit{space-time circuit}, e.g, $b_{x} \propto \int\int  T_{102} dt \wedge dy$. Here we took $e^{a}_{\mu}=\delta^{a}_{\mu}$, in both circumstances.

This is the above-mentioned geometric obstacle, that led earlier investigators to conclude that, for two dimensional Dirac materials, dislocations could not be accounted for by torsion \cite{deJuan2010}. Indeed, this is so if time is not considered. On the other hand, as we shall soon recall, in \cite{original_paper}, by focussing on the second scenario, it is shown that there might be a way out.



We can take the Riemann curvature to be zero, $\mathring{R}_{\mu \nu}^{ab}=0$, but with $\kappa_\mu^{ab}\neq0$, and choose a frame where $\mathring{\omega}_\mu^{ab}=0$ \cite{original_paper}. These settings make possible to isolate the effects of torsion on the system, and the corresponding action is
\begin{equation}\label{action_pure_torsion}
S = i\hbar v_F \int d^{3}x |e| \,  \left(\overline{\Psi}\gamma^{\mu}\partial_{\mu}\Psi - \frac{i}{4} \overline{\psi}_{+} \phi \psi_{+} + \frac{i}{4} \overline{\psi}_{-} \phi \psi_{-} \right) \;,
\end{equation}
where $\phi \equiv \frac{\epsilon^{\mu \nu \rho}}{|e|}T_{\mu\nu\rho}$. As clearly shown in (\ref{action_pure_torsion}), even in the presence of torsion, the two irreducible spinors, $\psi_{+}$ and $\psi_{-}$, actually decoupled. Nonetheless, they couple to $\phi$ with opposite signs.

\begin{figure}
\begin{center}
\includegraphics[width=0.25\textwidth,angle=90]{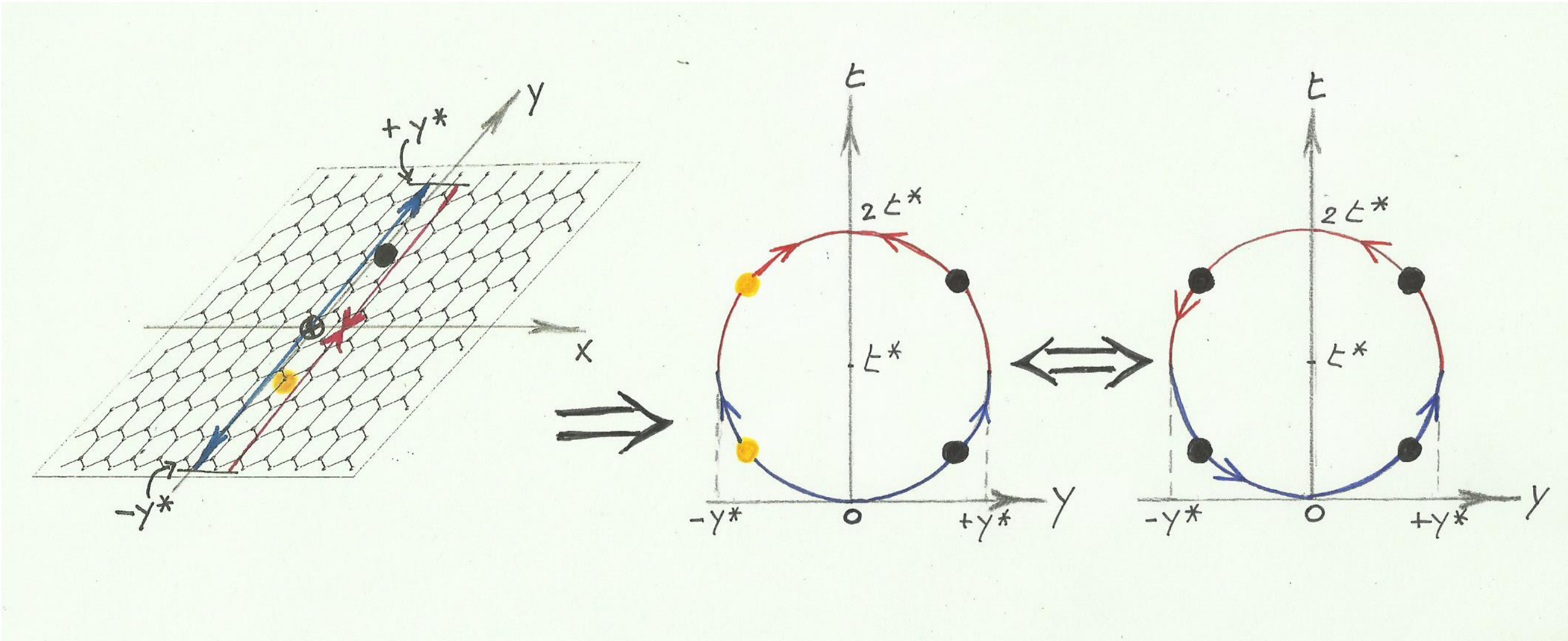}
\end{center}
\caption{Idealized \textit{time-loop}. At $t=0$, the hole (yellow) and the particle (black) start their journey from $y=0$, in opposite directions. Evolving forward in time, at $t=t^*>0$, the hole reaches $-y^*$, while the particle reaches $+y^*$,  (blue portion of the circuit). Then they come back to the original position, $y=0$, at $t=2t^*$ (red portion of the circuit). On the far right, the equivalent \textit{time-loop}, where the hole moving forward in time is replaced by a particle moving backward in time. Figure taken from \cite{original_paper}.}%
\label{Fig3TIMELOOP}%
\end{figure}

\begin{figure}
\begin{center}
\includegraphics[width=0.30\textwidth,angle=0]{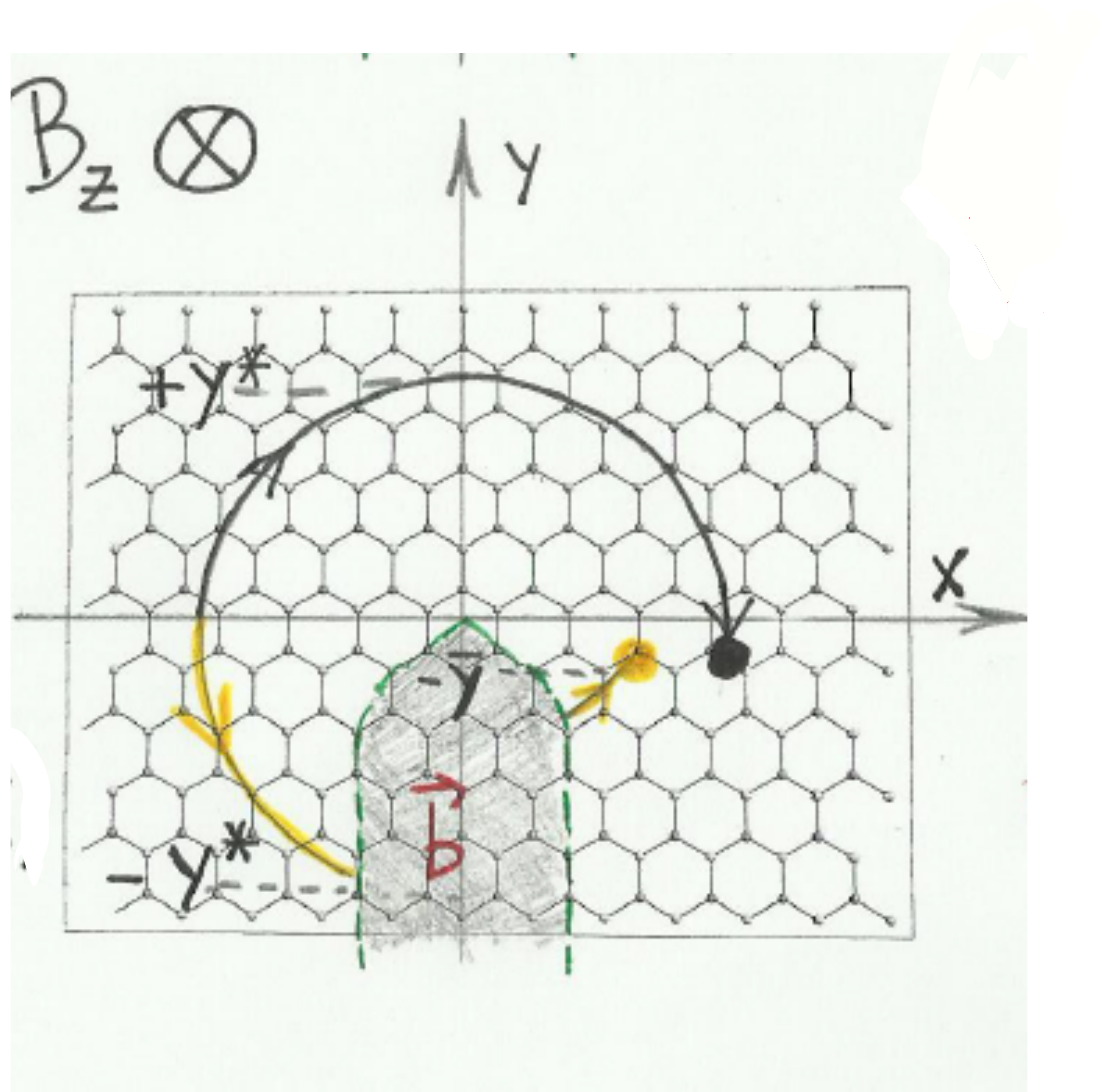}
\end{center}
\caption{Dislocation-induced deformations of the idealized \textit{time-loop}. Figure taken from \cite{original_paper}.}%
\label{Fig4NETFLUXandVERTEX}%
\end{figure}

To spot the effects of $\phi$, we propose to make use of the particle-antiparticle structure, encoded in the action (\ref{action_pure_torsion}). Indeed, the regime of Dirac materials we describe, is the ``half-filling'' \cite{WehlingDiracMaterials2014}, whose vacuum state has the vacancies of the valence band ($E < 0$) completely filled, and the vacancies of the conduction band ($E > 0$) empty. This is the analog of the Dirac sea, 
and the particle-antiparticle moving realizes a \textit{time-loop}. The pictures in Fig.~\ref{Fig3TIMELOOP} refer to a defect-free sheet. The presence of a dislocation, e.g., with Burgers vector $\vec{b}$ directed along $x$, would result in a failure to close the loop proportional to $\vec{b}$, as depicted in Fig.~\ref{Fig4NETFLUXandVERTEX}.

\section{Torsion in USUSY}
\label{section_USUSY}

In USUSY all the fields belong to a one-form connection $\mathbb{A}$, in (2+1) dimensions, and the vielbein is realized in a different way than in standard SUGRA models \cite{AVZ}. This model has nontrivial dynamics, and leads to a scenario where local SUSY is absent (although there is still diffeomorphism invariance), but rigid SUSY can survive for certain background geometries \cite{GPZ}. Because there is no local SUSY, there are neither SUSY pairings nor gauginos. The only propagating degrees of freedom are fermionic \cite{GPZ}, and the parameters that appear in the model are either dictated by gauge invariance, or raise as integration constants. The rigid USUSY invariant backgrounds are strong candidates to describe the $\pi$-electrons in graphene-like materials, see \cite{SU2Ususy}, and the more recent \cite{DauriaZanelli2019}. The versatility of USUSY make it possible to include non-Abelian internal groups, like $SU(2)$. This way one can take into account the two Dirac points at once, allowing for describing scenarios where, due to lattice symmetry considerations, both Dirac points are needed. This is the case of a continuous description of grain boundaries (a region in the lattice characterized by a misorientation angle between two sides). These appealing properties of USUSY were already reported in \cite{IORIO2018265}. Furthermore, in USUSY the torsion of geometric backgrounds appears very naturally, hence its totally antisymmetric part is coupled with the fermions. As discussed earlier, the third dimension to have a non-zero coupling is necessary, and this makes USUSY a good arena where to see the time-loop at work.

Taking into account the two Dirac points, the action of USUSY in $(2+1)$ dimensions for fixed background bosonic fields is obtained from the Chern-Simons three-form for $\mathbb{A}$ with an $SU(2)$ internal gauge group \cite{SU2Ususy} 
\begin{equation}\label{action_ususy}
S_{USUSY}=\kappa\int\overline{\psi}^{i}\left(\gamma^{\mu}\mathring{D}_{\mu}-\frac{i}{8}\tensor{\epsilon}{_{a}^{bc}}\tensor{T}{^{a}_{bc}}\right)\psi_{i}|e|d^{3}x\;,
\end{equation}
where lower case Latin letters, $a,b,\ldots$, represent tangent space Lorentz indices, and $\tensor{T}{^{a}_{bc}}=\tensor{T}{^{a}_{\mu\nu}}\,E^{\mu}_{b}\,E^{\nu}_{c}$. We are not taking into account possible boundary terms. Apart from a global factor $\kappa$ that can be adjusted to be $i\hbar v_{F}$, there are two main differences of $S_{USUSY}$ with respect to (\ref{action_torsion}). The first one is the coefficient in front of the torsion term, which appears in the system as an integration constant \cite{AVZ}. This difference is due to the coefficient associated to SUSY generators in $\mathbb{A}$ are composite fields (vielbein $+$ spin-$1/2$ fermion). The second difference is the index $i$ (here taken as a colour internal index, considering both Dirac points in the model), which is allowed by using another representation for $\psi$ and the Dirac matrices (see details in Appendix B of \cite{IORIO2018265}).


Finally, another attractive feature of USUSY is that it permits the description of a BTZ black hole \cite{BTZ}, in a pure bosonic vacuum state ($\psi=0$) \cite{AVZ}. This follows from the fact that the BTZ black hole can be obtained from a Lorentz-flat connection \cite{BTZ_Lorentz_flat}, provided the spacetime has torsion, in order that the contribution to Lorentz curvature coming from the contortion term cancels out the Riemann curvature contribution. The spectrum of BTZ black holes (as locally anti-de Sitter spaces, with negative cosmological constant $\Lambda=-1/\ell^{2}$), is given in terms of their mass, $M$, and angular momentum, $J$. This includes the extremal cases, $M\ell=|J|$ and $M=0$ (the $M=-1$ case is the globally anti-De Sitter space, while the other cases are conical singularities \cite{Miskovic2009}). In particular, the $M=0$ case could play a very important role in the gravity induced Generalized Uncertainty Principle \cite{Iorio_GUP_BTZ}, and in the related Hawking-Unruh phenomenon on graphene \cite{Iorio:2011yz}.

\section{Conclusions}
\label{section_conclusions}

When time is duly included in the emergent relativistic-like picture of Dirac materials, the geometric obstruction to describe the effects of dislocations in terms of a suitable coupling with torsion, within the $(2+1)-$dimensional field theoretical description of the $\pi$-electrons dynamics, in principle could be overcome. Provided dislocations can be meaningfully described by a suitable torsion tensor, the low energy Dirac field theory emerging here can include a nonzero coupling with torsion, accounting for a field theory description of the effects of dislocations, only when the third dimension is taken to be time. This also paves the road to the exploration of USUSY as a top-down description of the $\pi$ electrons of Dirac materials, where torsion appears in a very natural way.

\section*{Acknowledgments}
The authors thank Marcelo Ciappina and Adamantia Zampeli, for their collaboration, and Jorge Zanelli for inspiring discussions on these exciting matters. This research is partially supported through the grant Charles University Research Center (UNCE/SCI/013). P.~P. is also supported by by Fondecyt Grant No.~3200725.

\bibliographystyle{JHEP}
\bibliography{ICHEP_biblio}



\end{document}